\documentclass[aps,prl,twocolumn,showpacs,preprintnumbers,amsmath,amssymb,superscriptaddress]{revtex4}
\usepackage{graphicx}
\usepackage{dcolumn}
\usepackage{bm}

\begin{document}

\title{Interplay between folding and assembly of fibril-forming polypeptides}

\author{Ran Ni}
 \email{ran.ni@wur.nl}
 \affiliation{%
 Laboratory of Physical Chemistry and Colloid Science, Wageningen University, Dreijenplein 6, 6703 HB Wageningen, The Netherlands
 }%
 \affiliation{%
 Van $'$t Hoff Institute for Molecular Sciences, Universiteit van Amsterdam, Science Park 904, 1098 XH Amsterdam, The Netherlands
 }%

 \author{Sanne Abeln}
 \affiliation{%
Centre for Integrative Bioinformatics (IBIVU), Vrije Universiteit, De Boelelaan  1081A, 1081 HV Amsterdam, The Netherlands
 }%

\author{Marieke Schor}
 \affiliation{%
 Van $'$t Hoff Institute for Molecular Sciences, Universiteit van Amsterdam, Science Park 904, 1098 XH Amsterdam, The Netherlands
 }%

\author{Martien A. Cohen Stuart}
\affiliation{%
 Laboratory of Physical Chemistry and Colloid Science, Wageningen University, Dreijenplein 6, 6703 HB Wageningen, The Netherlands
 }%

\author{Peter G. Bolhuis}
\affiliation{%
Van $'$t Hoff Institute for Molecular Sciences, Universiteit van Amsterdam, Science Park 904, 1098 XH Amsterdam, The Netherlands
}%


\begin{abstract}
Polypeptides can self-assemble into  hierarchically organized fibrils consisting of a stack of individually folded polypeptides driven together by hydrophobic interaction.
Using a coarse grained model, we systematically studied this self-assembly 
as a function of temperature and  hydrophobicity of the residues on the outside of the building block.
We find the self-assembly can occur via two different pathways --- a random aggregation-folding route, and a templated-folding process ---  thus indicating a strong  coupling between folding and assembly.  
The simulation results can explain experimental evidence that assembly through stacking of folded building blocks is rarely observed, at the experimental concentrations. The model thus provides a generic picture of hierarchical fibril formation.
\end{abstract}



\maketitle

While, perhaps surprisingly, most proteins, and even short peptides, share the general ability to form fibrils under appropriate conditions~\cite{fib1,fib2,fib3}, the interest in filamentous proteins originates to a large extent from their association with neurodegenerative disorders such as Alzheimer’s and Parkinson’s disease~\cite{disease}. However,  protein fibers also have promising applications in biomaterials~\cite{bioapp1,bioapp2}. For instance, 
silk-collagen-like triblock copolymers self-assemble into micrometer long fibrils, which form dilute gels with surprisingly high modulus, serving as promising candidates for novel materials such as artificial tissues~\cite{fibexp}. 
Similar to  amyloid fibrils, these silk-collagen fibrils consist of stacked cross $\beta$ structures made out of the silk-like block, sterically stabilized by the collagen-like hydrophilic flanks.  Contrary to amyloid fibrils, the silk-like building blocks are so-called  $\beta$-rolls, consisting of two interconnected parallel $\beta$-sheets with a hydrophobic outside surface, which promotes stacking into long fibers~\cite{schor2009}. 
The mechanism of such hierarchical protein fibril formation  is still poorly understood. Yet, such understanding is crucial for controlling fibril nucleation and growth. 
One clue comes from  the fact that experimental fibril elongation speeds are more than four orders of magnitude smaller than theoretically predicted by diffusion~\cite{lennart2012}. 
This suggests that the fibril growth (and nucleation) encounters pronounced free energy barriers, possibly due to significant conformational changes, e.g., folding, of each polypeptide. Hierarchical fibril formation thus involves a combination of both folding and assembly processes. One can therefore imagine (at least) three possible
scenarios:  (i) an aggregation-folding process, in which from disordered aggregates of unfolded polypeptides a fibril structure nucleates and grows, (ii) a templated-folding process, in which existing fibrils sequester unfolded polypeptides and induce them to fold, or (iii) a folding-docking process, in which polypeptides first fold individually before stacking together into fibrils.
Computer simulation can complement experiments and yield mechanistic insight, but due to computational challenges, the folding and assembly processes of fibril-forming polypeptides are usually studied separately~\cite{aggfibril,auerphasediag,review,schor2009,schor2011,ckhall0,ckhall1,ckhall2}.  Thus, the interplay between folding and assembly in the self-assembly of polypeptides is still an open question.
Here, we investigate such  interplay for a system closely related to the silk-collagen-like triblock copolymer, which is, due to its hierarchical structure, an excellent model system. 

While in principle atomistic models can provide molecular insight in the folding and assembly process, such all-atom simulations are prohibitively expensive. 
Instead we employ a coarse-grained polypeptide model in which each residue occupies a single site on a  cubic lattice~\cite{sanne2011}, and all other sites are considered as solvent.  For each residue  a unit vector indicates the direction of the side chain.
This highly efficient model captures two  essential features for a correct description of folding: the formation of hydrogen bonds and the directionality of side chains.
The total potential energy of the system is given by
$	E = E_{aa} + E_{solvent} + E_{hb} + E_{steric},$
where $E_{aa}$ and $E_{solvent}$ are a knowledge based interaction potential between amino acids, and between 
amino acids and solvent, respectively (see Table I). This potential has been shown to prevent artificial aggregation of proteins in their native state.
$E_{hb}$ is the potential energy of formed hydrogen bonds, and $E_{steric}$ represents the steric hindrance between consecutive residues in a polypeptide chain~\cite{sanne2013}. 
Two amino acids in contact interact only when their side chains are either in parallel or pointing towards each other. 
Similarly, interaction between a residue and solvent only exists when 
the side chain points to a solvent site. 
 When a residue is not part of a turn in the backbone, it can adopt a {\it strand} state, depending on the side chains. Two residues in contact and both in strand state can  form a hydrogen bond with an energy $\epsilon_{hb} = -0.5$ (all interaction potentials in this work are in reduced units), when their side chains are aligned.
An energy penalty, $\epsilon_{s} = 0.55$,  prevents that side chains of consecutive residues point in the same direction, mimicking steric hindrance and restrictions in bond rotation.
Due to its small size alanine becomes considerably more hydrophobic in a $\beta$-strand environment\cite{suppinfo}. This effect is not captured in the original parameterization of the potential. To compensate for this, and to investigate the influence of the hydrophobicity of the building block's surface, the interaction between alanine (A) and water (w) is varied between 0.01 and 0.6 (see Table I). 
The model is  simulated using lattice Monte Carlo with a classic move set~\cite{sanne2011}.

\begin{table}[t]
\caption{Interaction matrix (in reduced units) for the residues in the used sequence~\cite{sanne2011}. Amino acids are denoted by their one-letter code (I = isoleucine, A = alanine, R = arginine, E = glutamate) and $w$ denotes the solvent (water). The hydrophobicity of alanine is varied via the A-w interaction ($\epsilon_{A,w}$) as indicated.}\label{intermatrix}
\begin{tabular}{rrrrrr} 
 	& I 		& A 		& R 		& E 		& w\\
\hline
 I	& -0.79	& -0.40	& 0.5	& 0.69	& 0.7\\
 A	&		& -0.34	& 0.49	& 0.77	& 0.01; 0.1; 0.2; 0.4; 0.6\\
 R	&		&		& 0.43	& -0.6	& -0.57\\
 E	&		&		&		& 1.02	& -0.78\\
 w	&		&		&		&		& 0.0\\
 \hline
\end{tabular}
\end{table}

\begin{figure}[t]
	\includegraphics[width=0.3\textwidth]{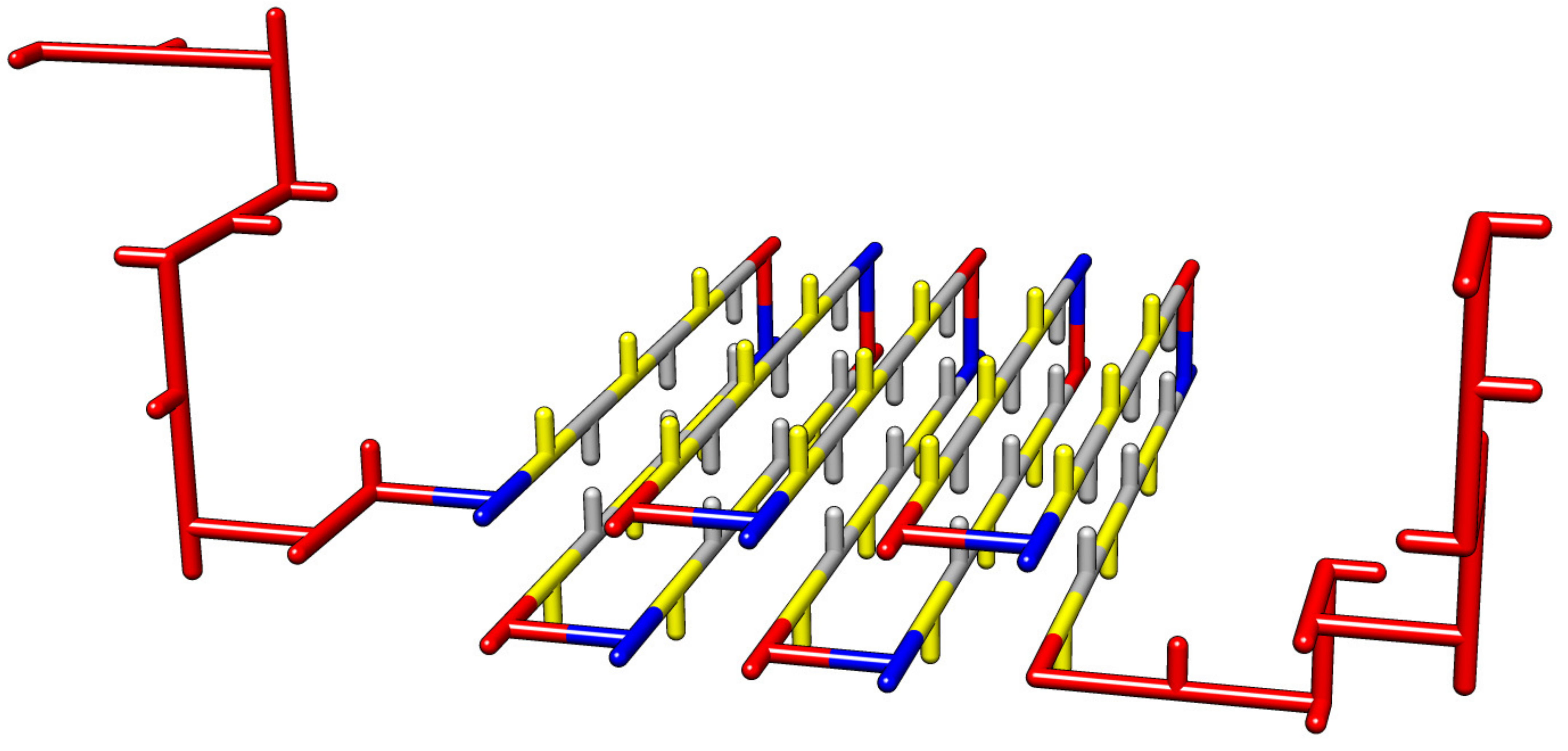}
	\includegraphics[width=0.45\textwidth]{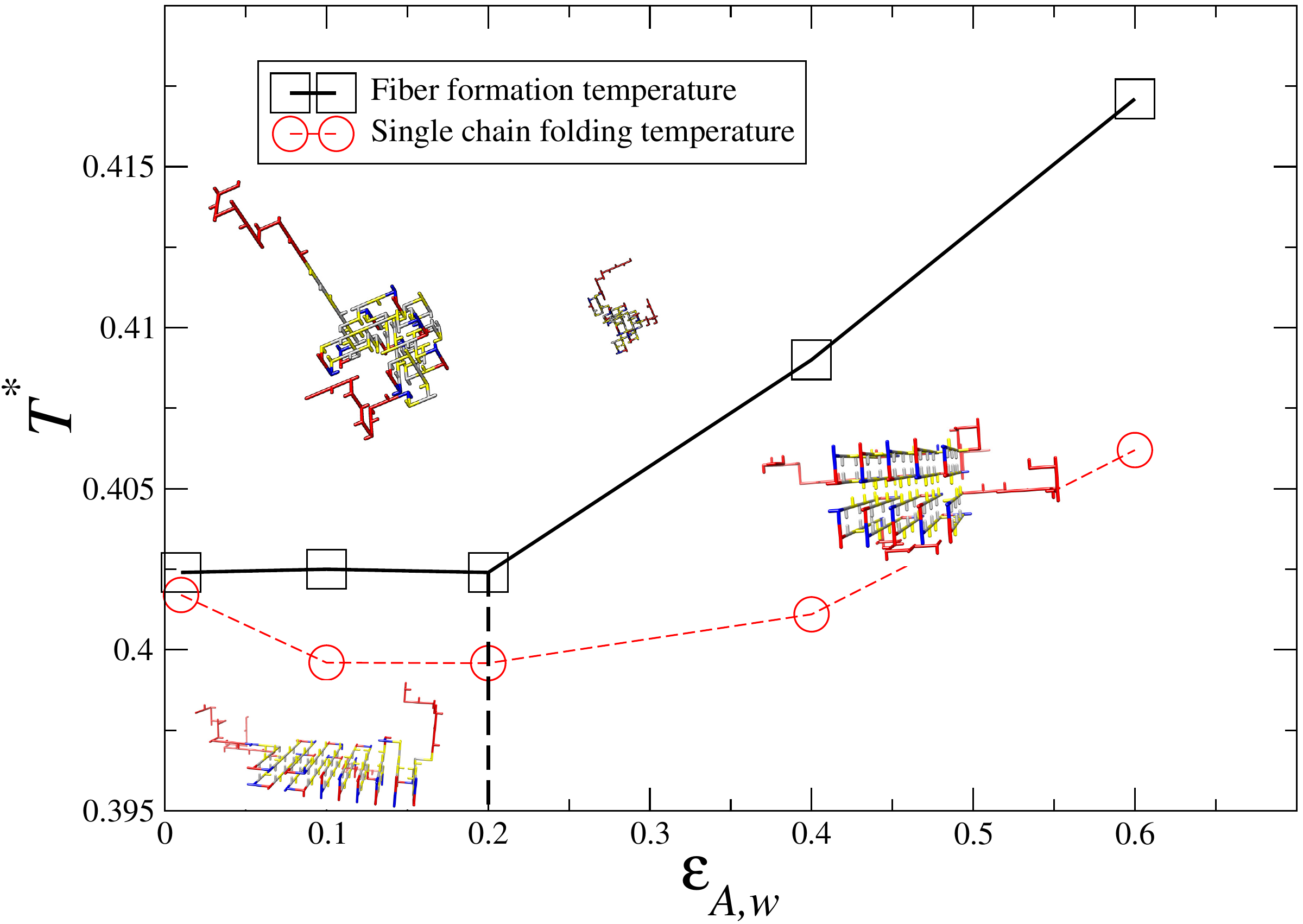}
	\caption{(color online) Top: Folded $\beta$-roll structure of a single polypeptide (alanine(A)=yellow, isoleucine(I)=white, glutamate(E)=red, arginine(R)=blue). Bottom: Equilibrium phase diagram in the $T^*- \epsilon_{A,w}$ plane, with transition temperatures  for two polypeptides (squares) and the folding temperature of single polypeptide (circles). 
Typical snapshots of the isolated unfolded, aligned and stacked phases are included.}\label{fig2}
\end{figure}

Lattice models in general are not expected to fold natural sequences. We therefore first need to design a sequence to fold into the desired $\beta$-roll structure. Proper  folding of a  $\beta$-roll on a  cubic lattice demands a palindromic sequence and a slightly different topology compared to its off-lattice counterpart: whereas
the experimental  $\beta$-roll consists of two interconnected parallel $\beta$-sheets, the lattice $\beta$-roll is comprised of anti-parallel sheets. 
Restricting the silk part to 80 residues to make the calculation tractable,
the design procedure results in the optimal sequence $(E(AI)_3RE(IA)_3R)_5$. This sequence has a repetitive nature comparable to the experimental silk block $(E(AG)_3))_{24}$ \cite{lennart2012}, but with 
isoleucine replacing  glycine. This is not unrealistic, as glycines in $\beta$-sheets are more hydrophobic than the average glycine~\cite{suppinfo}.  
 The extra (arginine) residue has been introduced to fit the $\beta$-roll structure on the lattice. 
In experiments, random aggregation of fibrils is prevented by the presence of hydrophilic flanks\cite{lennart2012}. A similar effect has been shown  by lattice model simulations \cite{abeln2008}. 
To mimic those flanks we attached 10 glutamate residues to both peptide termini. 
As shown in Fig.~\ref{fig2}, the designed polypeptide indeed folds into a $\beta$-roll structure with a hydrophobic outside surface covered with  alanine side-chains.

Whereas polypeptide folding is driven by several types of interactions, including hydrophobicity and hydrogen bond formation, the assembly is purely hydrophobicity driven\cite{suppinfo}.
To investigate the interplay between folding and assembly, it is necessary to be able to influence both independently. Control parameters in the folding and assembly process are the temperature and the concentration, but these will probably influence both aspects simultaneously. Therefore we also change 
the hydrophobic strength of the outside surface, by varying the alanine-solvent interaction.

We performed replica exchange Monte Carlo (REMC) simulations of the optimized sequence
on a $200^3$  lattice with periodic boundary conditions. Each REMC simulation consisted of 64 replicas with a (reduced) temperature distribution around the transition temperature ($0.35 \le T^* \le 0.45$), optimized by a feedback-optimization algorithm ~\cite{fbo}.  A replica exchange was attempted each 1000 MC cycles.
After equilibrating for  $1 \times 10^{10}$ MC cycles per replica, we  performed
$5 \times 10^{10}$ MC cycles  for production. Employing the virtual-move parallel tempering method~\cite{vmpt} ensured optimal use of  the simulation data.

\begin{figure*}[t]
	\includegraphics[width=0.4\textwidth]{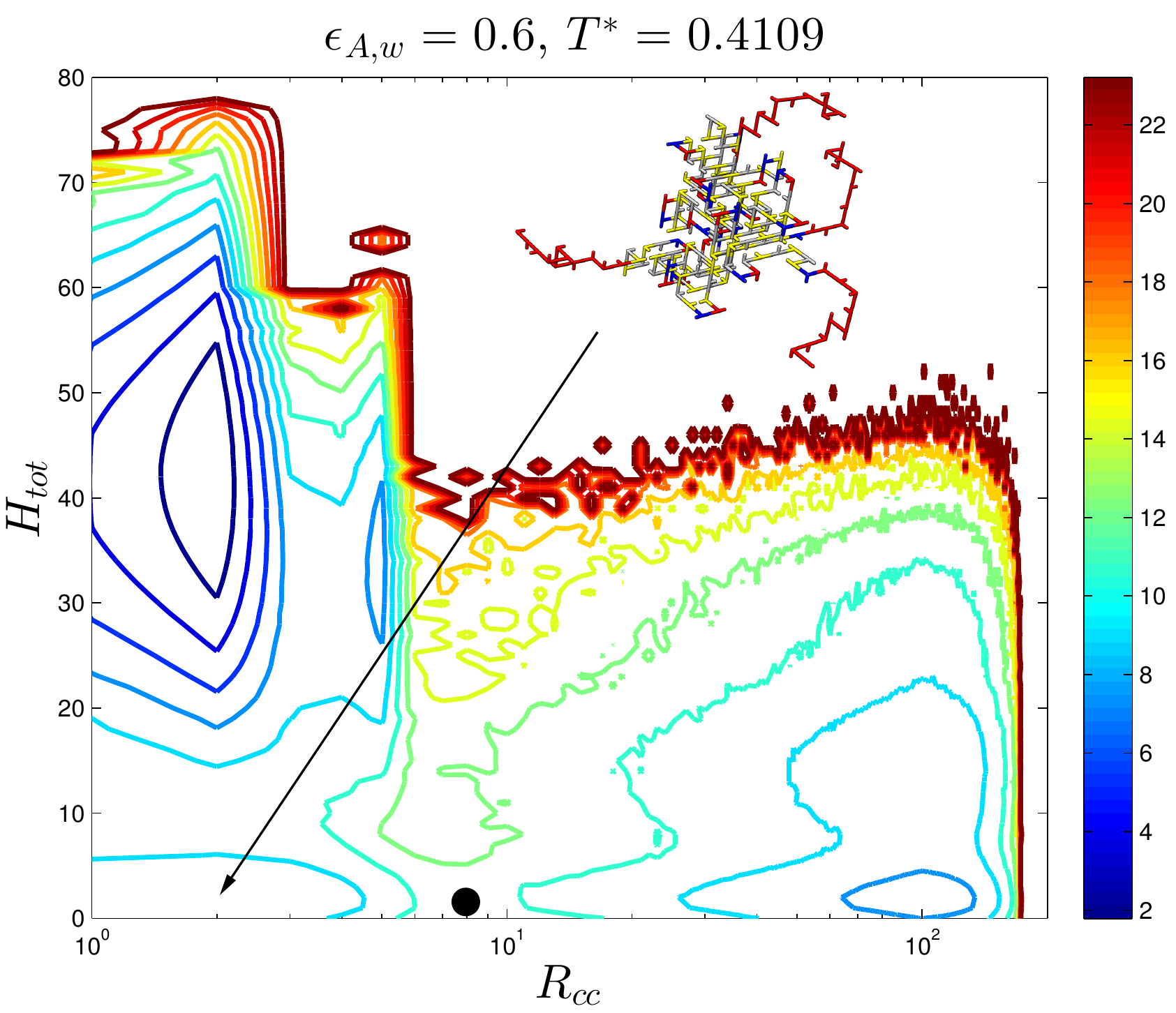}	\includegraphics[width=0.4\textwidth]{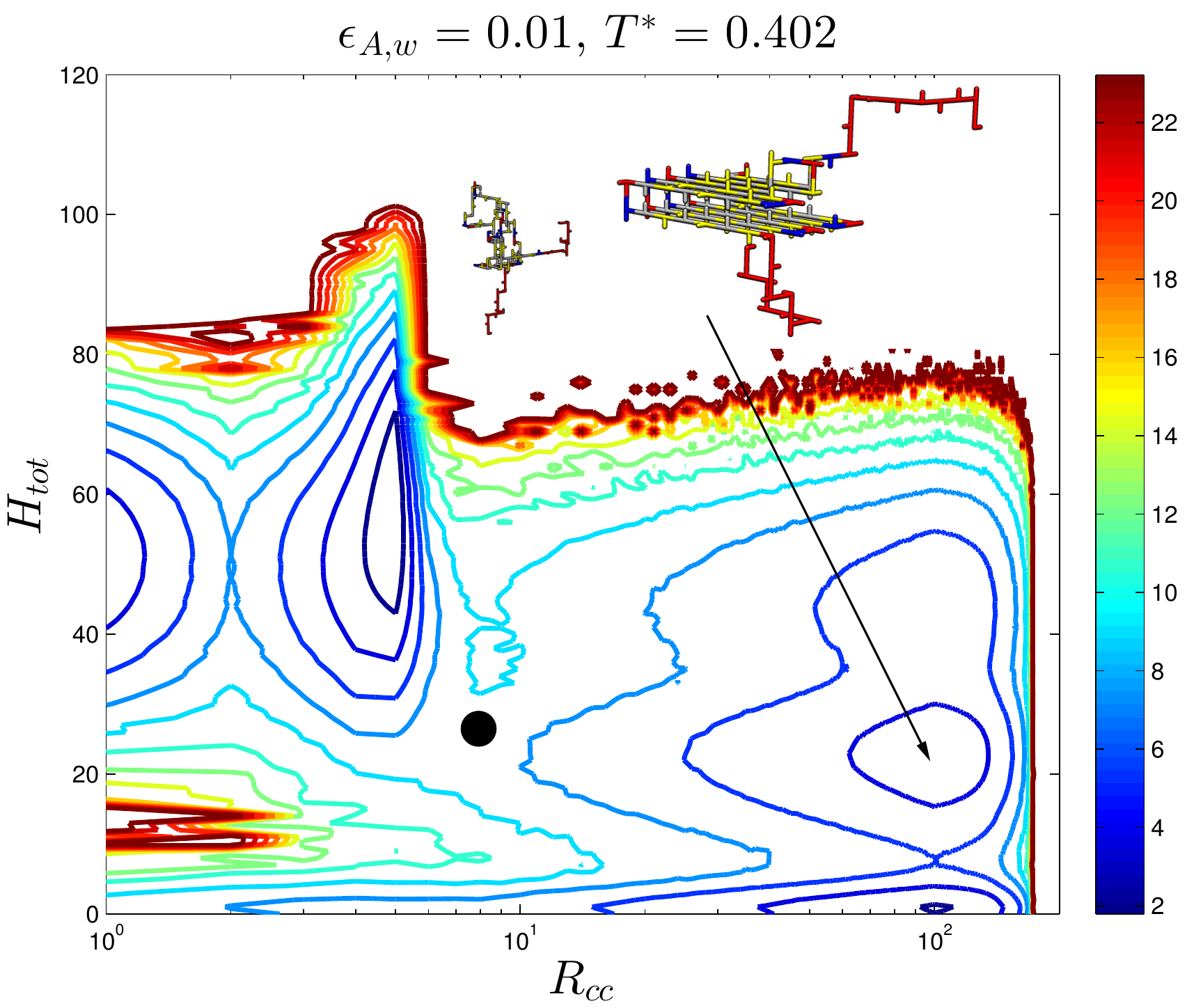}\\
	\includegraphics[width=0.4\textwidth]{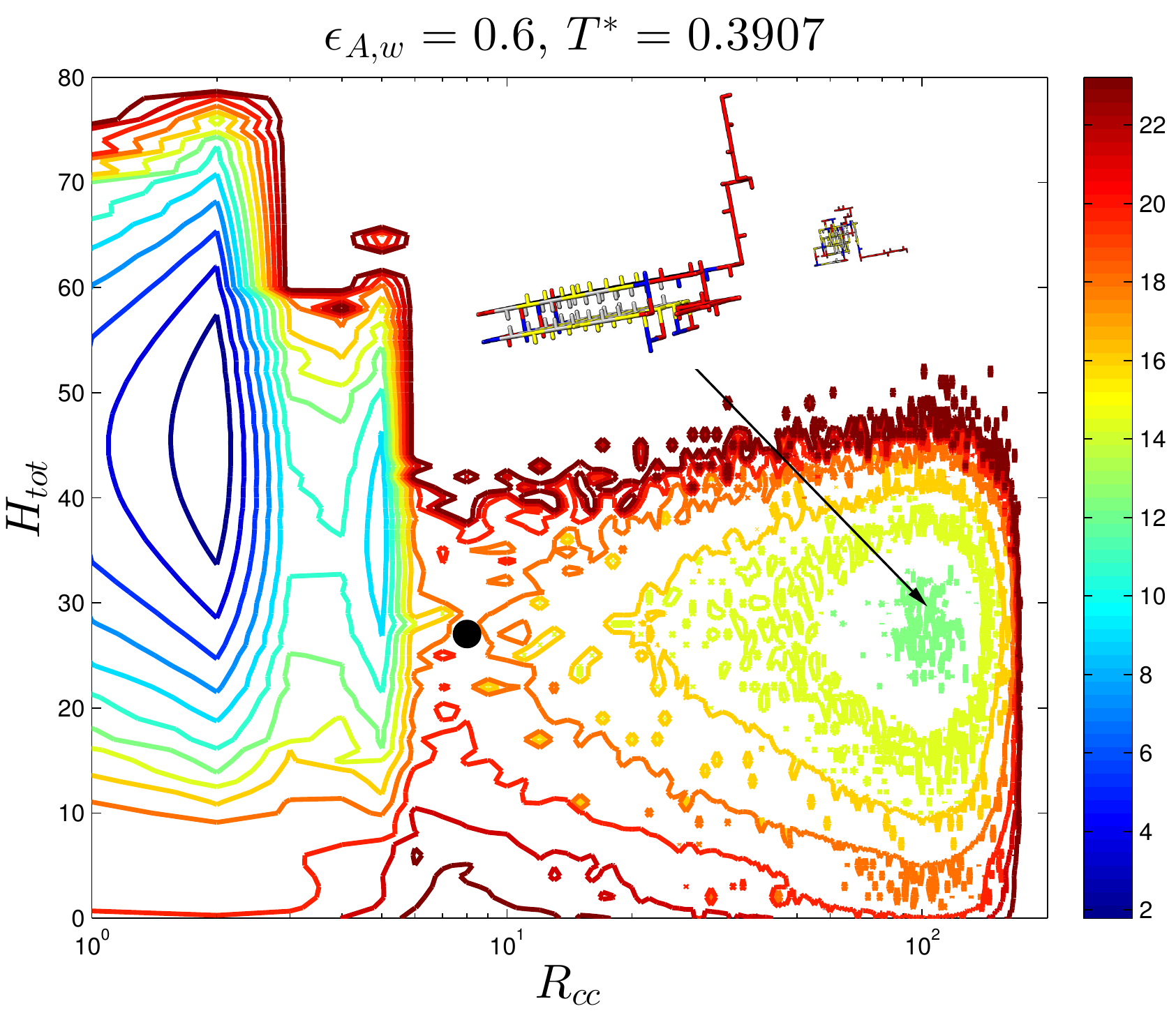} 	\includegraphics[width=0.4\textwidth]{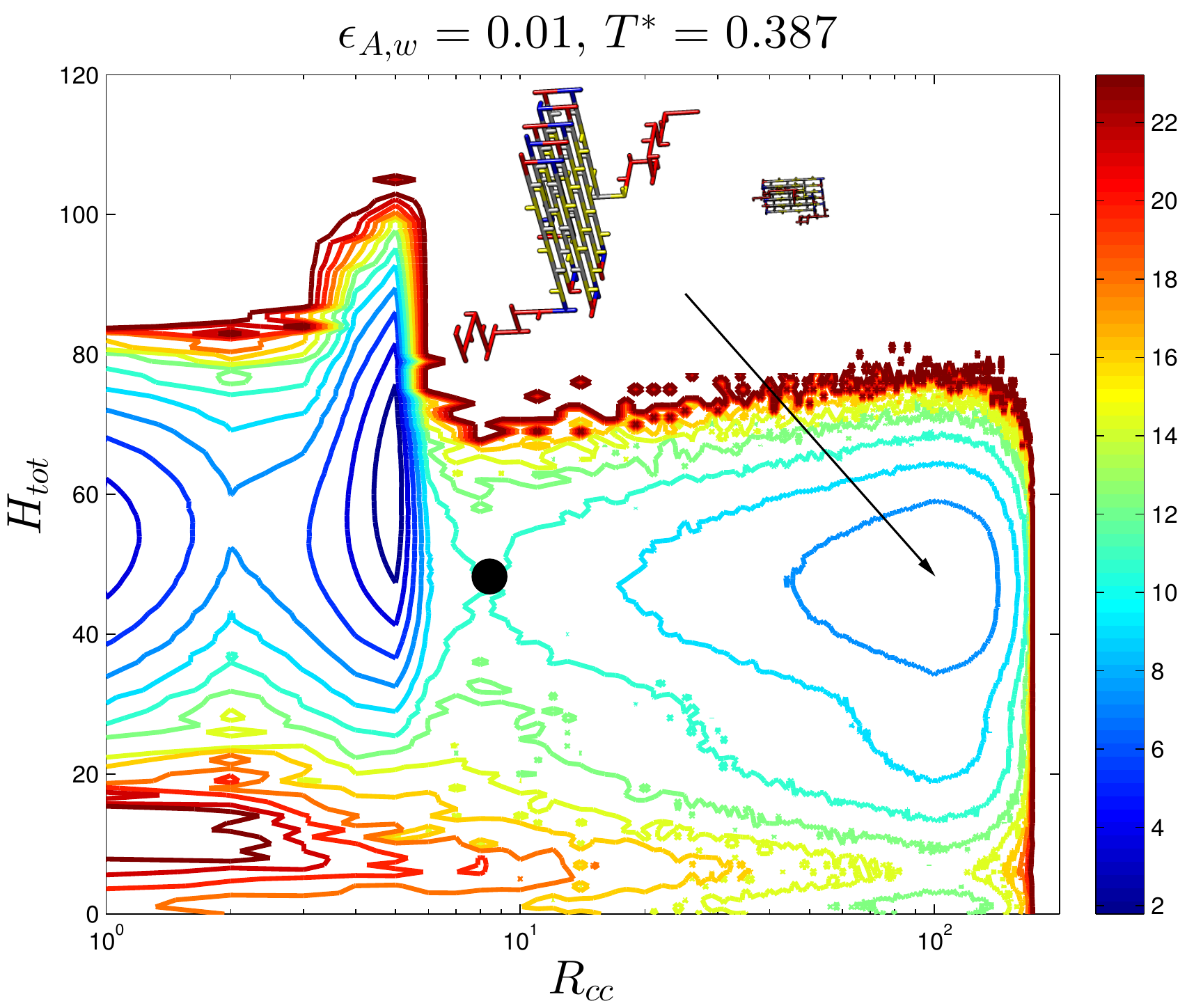}\\
	\caption{(color online) Free energy landscape as a function of the silk-block center-to-center distance $R_{cc}$, and the total number of hydrogen bonds, $H_{tot}$. Left column: $\epsilon_{A,w} = 0.6$,  $T^* = 0.4109$ (top) and 0.3907 (bottom). Right column: $\epsilon_{A,w} = 0.01$ at the temperature $T^* = 0.402$ (top) and 0.387 (bottom). Solid circles denote saddle points.}\label{fig3}
\end{figure*}

For a single polypeptide chain, corresponding to infinite dilution, we obtained the folding temperature
from the heat capacity peak.  Fig.~\ref{fig2} shows that this temperature is weakly dependent on the alanine hydrophobicity; at high $\epsilon_{A,w}$ the folded chain is slightly more stable, possibly because  alanine side chains are more shielded from the water.
In addition, with increasing hydrophobicity we observe more collapsed structures in the unfolded state. 

Next, we performed REMC simulations of two polypeptides.
Assuming a lattice space of 0.38nm~\cite{sanne2011}, this corresponds to a concentration of $7.57 \mu \mathrm{M}$. 
At high temperature the polypeptides are unfolded and  
mostly separated (see top left of Fig.~\ref{fig2}). Upon decreasing temperature, a sharp peak in the heat capacity  signals a transition to a self-assembled folded state, with a structure that 
depends on the  alanine hydrophobicity $\epsilon_{A,w}$. 
For $\epsilon_{A,w} < 0.2$, the two folded $\beta$-rolls align side-by-side.  As in this regime the transition temperature  remains almost constant, the formation of aligned state is not driven by the hydrophobicity of the $\beta$-roll surface, but instead by hydrogen bonds between the sides of  two $\beta$-rolls.
For  $\epsilon_{A,w} \ge 0.2$, the two folded $\beta$-rolls are stacked on top of each other. In this regime the transition temperature increases monotonically with $\epsilon_{A,w}$ 
as the stacked state is stabilized by the hydrophobicity of the
outside surface of the folded $\beta$-roll.

To shed light on the self-assembly mechanism of fibril-forming polypeptides, we plot in  Fig.~\ref{fig3}(left), 
for two temperatures, 
the free energy landscapes associated with the stacked state formation at $\epsilon_{A,w} = 0.6$ as a function of the total number of hydrogen bonds, $H_{tot}$, and the silk-block center-to-center distance $R_{cc}$. 
For both temperatures  the global free energy minimum at $(R_{cc} = 2,H_{tot} \simeq 45)$  corresponds to the stacked state, and a local minimum  at $(R_{cc} \simeq 5,H_{tot} \simeq 35)$ represents the metastable aligned state. 
However, the mechanistic folding pathways 
differ significantly for the two temperatures.  At $T^* = 0.4109$, above  the single polypeptide  folding temperature, 
there exists (besides the minimum at $(R_{cc} \simeq 100,H_{tot} \simeq 0)$ corresponding to isolated random coils) a second local minimum at $(R_{cc} \simeq 2,H_{tot} \simeq 0)$, associated with a disordered aggregate of two polypeptides.
The  presence of this aggregated state causes the folding mechanism  
to pass through a saddle point around $(R_{cc} \simeq 8,H_{tot} \simeq 1)$. 
This implies that above the single polypeptide  folding temperature  the self-assembly follows the  pathway: unfolded state $\rightarrow$ random aggregation $\rightarrow$ stacked state. 

At $T^* = 0.3907$, below the single polypeptide folding temperature, the free energy landscape changes significantly, and features a local minimum around $(R_{cc} \simeq 100,H_{tot} \simeq 30)$, which corresponds to an intermediate state consisting of  a separated folded and unfolded polypeptide. The channel connecting this  state to the stacked state travels through a saddle point around $(R_{cc} \simeq 10,H_{tot} \simeq 30)$, suggesting {that} the assembly is a self-templated process~\cite{auer2008}, in which  a polypeptide folds on top of an already folded polypeptide. At very low temperatures the two polypeptides may 
fold  independently before stacking together. 
However, we failed to observe such folding-docking process  down to $T^* = 0.35$, more than $15\%$ below the transition temperature. Apparently, this is a very unlikely process, which explains why in experiments
individual $\beta$-rolls are rarely observed, whereas they readily form on the growing end of an existing filament~\cite{lennart2012}.

For an alanine hydrophobicity of  $\epsilon_{A,w} =0.01$ typical free energy landscapes in Fig.~\ref{fig3}({right}) show that below the transition temperature, the global minimum 
 is located at $(R_{cc} \simeq 5, H_{tot} \simeq 60)$, corresponding to  the aligned state (see  Fig.~\ref{fig2}). For $T^* = 0.402$, just below the transition temperature,  the assembly already occurs via the self-templated process, as indicated by the saddle point at $(R_{cc} \simeq 8, H_{tot} \simeq 25)$ and the metastable intermediate state located at $(R_{cc} \simeq 100, H_{tot} \simeq 25)$. The transition temperature is very close to the single polypeptide folding temperature (see Fig.~\ref{fig2}), revealing that the binding between the folded polypeptides is weak. At $T^* = 0.387$ one finds individual folded polypeptides in the solution, as indicated by a local minimum at $(R_{cc} \simeq 100, H_{tot} \simeq 50)$. 
The saddle point at $(R_{cc} \simeq 10, H_{tot} \simeq 50)$ suggests that the assembly then occurs via the folding-docking process. The  local minimum at $(R_{cc} \simeq 1, H_{tot} \simeq 50)$  corresponds to a metastable structure consisting of two interlocked folded $\beta$-sheets, which is probably an artifact of the model.
In contrast to the strong hydrophobicity case, there is no third pathway via a random aggregate. 
We note that the high hydrophobicity case is probably a more realistic approximation to the experiments of Ref.~\cite{lennart2012}.

\begin{figure}[t]
	\includegraphics[width=0.45\textwidth]{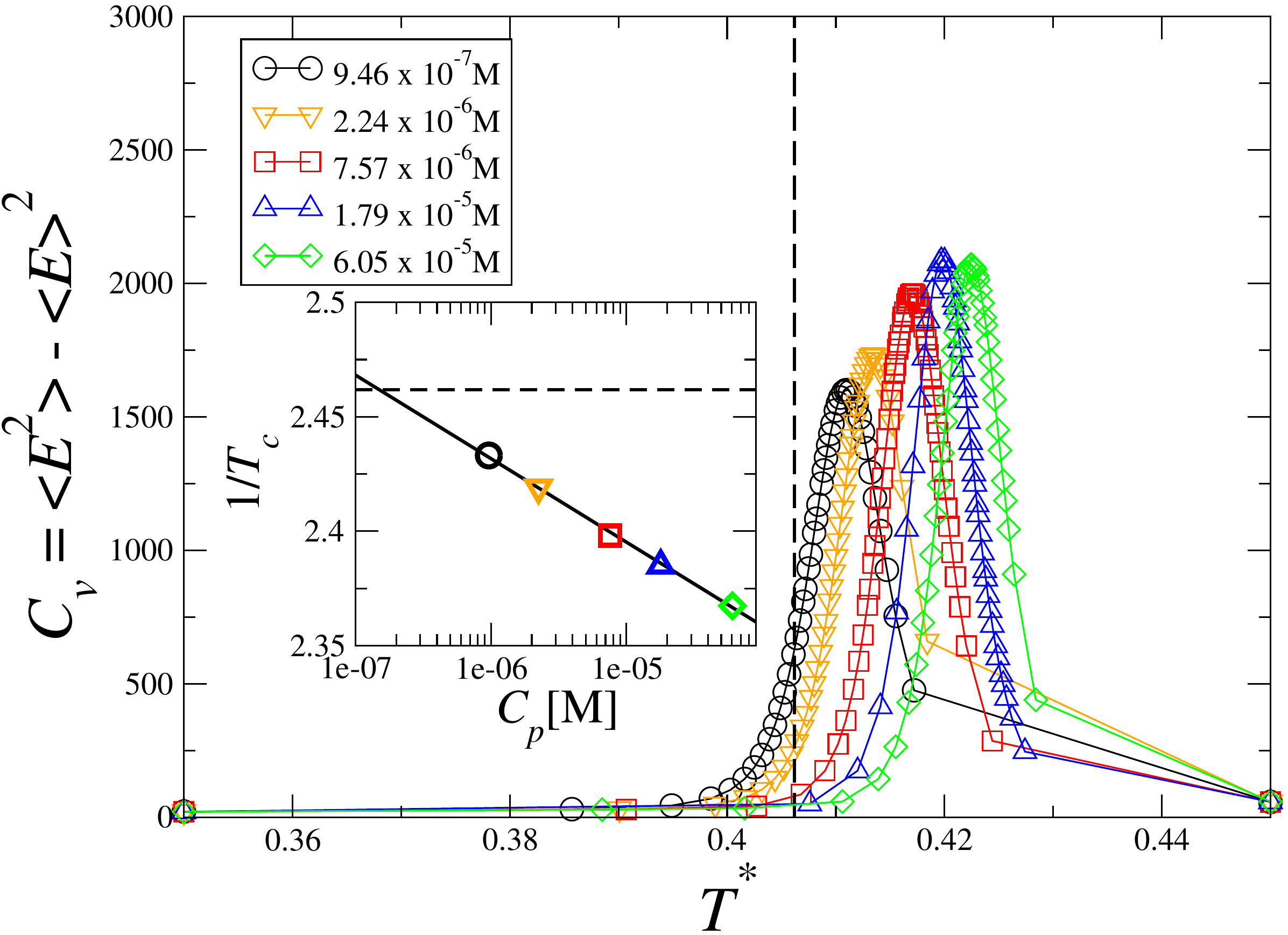}
	\caption{(color online) Main: Heat capacity versus temperature for two polypeptides at various concentrations with $\epsilon_{A,w} = 0.6$.
Inset: 
The reciprocal transition temperature $T_c$ versus the logarithm of peptide concentration. In both plots the  dashed line indicates the single polypeptide folding temperature. }\label{fig4}
\end{figure}

To investigate the effect of  polypeptide concentration, we performed REMC simulations at concentrations ranging from 1 to $60\mu \mathrm{M}$, by changing the  lattice size. Figure~\ref{fig4}  shows the heat capacity for $\epsilon_{A,w} = 0.6$  as a function of temperature.
The decreasing height of the  heat capacity peak with decreasing concentration  indicates the self-assembly process becomes less cooperative  at lower concentration. Still, the self-assembly transition temperature is always above the single polypeptide  folding temperature, and the mechanism does not change. The inset in Fig.~\ref{fig4} shows the linear relationship between the reciprocal transition temperature and the logarithm of polypeptide concentration. 
We explain this behavior as follows.
If assembly of two peptides  yields an energy gain $\Delta E<0$, then at the transition temperature $T_c$ the  free energy difference between the isolated and assembled state is $\Delta F = \Delta E - T_c  \Delta S = 0$, where $\Delta S$ is the entropy lost in the assembly. Assuming ideal gas behavior {of isolated peptides} $ \Delta S =  k_B \ln C_p - \Delta S_c$, {where} $C_p$ is the polypeptide concentration. The first term is the translational entropy lost, while the second term is the conformational entropy lost $(\Delta S_c> 0)$. It follows that  $1/T_c = (k_B/\Delta E) \ln C_p  - \Delta S_c/\Delta E = a_1 \ln C_p + a_2$, with $a_1<0$ and $a_2>0$, as we find in the simulations. 
We note that at very low concentration $(C_p < 0.1 \mu M)$ the assembly will not influence the folding temperature anymore, as indicated by the dashed line in the inset.
We speculate that below this concentration a folding-docking mechanism may indeed be present.
For very high concentration $C_p \sim 1/8R_g^3$, with the $R_g$ the radius of gyration, the unfolded chains are touching, and $T_c$ will also become independent of concentration.

In conclusion, we used a coarse-grained lattice model to systematically study the interplay between folding and assembly in the formation of hierarchically organized polypeptide fibrils. Despite the simplicity of the model, the results not only show a remarkably rich phase behaviour, but also, and most importantly, reveal a generic mechanism for protein fibril formation, depending on the relative strength of the intermolecular driving force for assembly (the hydrophobicity of the folded protein surface) with respect to the strength of the intramolecular driving force for protein folding (the hydrogen bonding). Coupling between these driving forces leads to an effect similar to \textit{allostery} where a conformation change goes along with binding, and produces a templated-folding growth process at temperatures below the melting point. The model explains why experiments rarely detect the presence of individual $\beta$-rolls, but instead provide evidence for a templating mechanism for nucleation and growth~\cite{lennart2012}.

Our results show that  the formation and growth  of fibril structures can be tuned by varying  the temperature and/or the hydrophobicity of specific residues,  and  are thus not only of fundamental interest, but may also provide guidance to fabricating novel bio-fibrils with high quality in experiments.

\begin{acknowledgments}
	We thank Erik van Dijk of Vrije Universiteit Amsterdam for helpful discussions. We acknowledge financial support from the European Research Council through Advanced Grant 267254 (BioMate). 
This work is part of the research programme VICI 700.58.442, which is financed by the Netherlands Organization for Scientific Research (NWO). SA has been supported by a Veni grant on the project 'Understanding toxic protein oligomers through ensemble characteristics' from Netherlands Organization for Scientific Research (NWO).
\end{acknowledgments}

\bibliography{ref}

\end{document}